\documentclass[journal,twocolumn,10pt,twoside]{IEEEtran}
\normalsize

\ifCLASSINFOpdf
   \usepackage[pdftex]{graphicx}
   \usepackage{epstopdf}
\else
\fi

\usepackage{amsmath}
\usepackage{amssymb}
\usepackage{cite}
\usepackage{graphicx}
\usepackage{algorithm}
\usepackage{algpseudocode}
\usepackage{subfigure}
\usepackage{multirow} 
\usepackage{longtable}
\usepackage{threeparttable}
\usepackage{caption3}
\usepackage{array}
\usepackage{cases}
\usepackage{color}
\usepackage[T1]{fontenc}
\usepackage[utf8]{inputenc}
\usepackage{authblk}
\usepackage{subfigmat}

\newtheorem{remark}{Remark}

\begin{document}
\title{Rate-Splitting Unifying SDMA, OMA, NOMA, \\ 
and Multicasting in MISO Broadcast Channel: \\
A Simple Two-User Rate Analysis}

\author{
\IEEEauthorblockN{Bruno Clerckx, Yijie Mao, Robert Schober, and H. Vincent Poor} 
\vspace{-10mm} 
\thanks{B. Clerckx is with Imperial College London, London SW7 2AZ, UK (email: b.clerckx@imperial.ac.uk). Y. Mao is with The University of Hong Kong, Hong Kong, China (email: maoyijie@eee.hku.hk). R. Schober is with University
of Erlangen-Nuremberg, 91058 Erlangen, Germany (email: robert.schober@fau.de). H. V. Poor is with Princeton University, Princeton, NJ 08544 USA (e-mail: poor@princeton.edu). This work has been partially supported by the EPSRC of the UK under grant EP/N015312/1.
}
}

\maketitle

\begin{abstract} Considering a two-user multi-antenna Broadcast Channel, this paper shows that linearly precoded Rate-Splitting (RS) with Successive Interference Cancellation (SIC) receivers is a flexible framework for non-orthogonal transmission that generalizes, and subsumes as special cases, four seemingly different strategies, namely Space Division Multiple Access (SDMA) based on linear precoding, Orthogonal Multiple Access (OMA), Non-Orthogonal Multiple Access (NOMA) based on linearly precoded superposition coding with SIC, and physical-layer multicasting. The paper studies the sum-rate and shows analytically how RS unifies, outperforms, and specializes to SDMA, OMA, NOMA, and multicasting as a function of the disparity of the channel strengths and the angle between the user channel directions. 

\end{abstract}
\begin{IEEEkeywords} Rate-splitting, multi-antenna broadcast channel, rate analysis, SDMA, OMA, NOMA, multicasting
\end{IEEEkeywords}

\IEEEpeerreviewmaketitle
\vspace{-0.4cm}
\section{Introduction}
\par Linearly precoded Rate-Splitting (RS) with Successive Interference Cancellation (SIC) receivers has recently appeared as a powerful non-orthogonal transmission and robust interference management strategy for multi-antenna wireless networks \cite{Clerckx:2016}. 
Though originally introduced for the two-user Single-Input Single-Output Interference Channel (IC) in \cite{Han:1981}, RS has become an underpinning communication-theoretic strategy to tackle modern interference-related problems and has recently been successfully investigated in several Multiple-Input Single-Output (MISO) Broadcast Channel (BC) settings, namely, unicast-only transmission with perfect Channel State Information at the Transmitter (CSIT) \cite{Mao:2017,Mao:2018b} and imperfect CSIT \cite{Yang:2013,Hao:2015,Joudeh:2016a,Joudeh:2016b,Dai:2016,Davoodi:2016,Piovano:2017,Dai:2017,Lu:2018}, (multigroup) multicast-only transmission \cite{Joudeh:2017}, as well as superimposed unicast and multicast transmission \cite{Mao:2018}. Results highlight that RS provides significant benefits in terms of spectral efficiency \cite{Mao:2017,Hao:2015,Joudeh:2016a,Dai:2016,Lu:2018,Joudeh:2017,Mao:2018}, energy efficiency \cite{Mao:2018b}, robustness \cite{Joudeh:2016b}, and CSI feedback overhead reduction \cite{Hao:2015,Dai:2017} over conventional strategies used in LTE-A/5G that rely on fully treating interference as noise (e.g. conventional multi-user linear precoding and Space Division Multiple Access - SDMA) or fully decoding interference (e.g. power-domain Non-Orthogonal Multiple Access - NOMA \cite{Liu:2017}). The key behind realizing those benefits is the ability of RS, through splitting messages into common and private parts, to partially decode interference and partially treat interference as noise. %This enables RS to softly bridge and therefore reconcile the two extreme strategies of fully decode interference and treat interference as noise. 
Additionally, RS is an enabler for powerful multiple access designs that subsumes SDMA and NOMA as special cases and outperforms them both for a wide range of network loads (underloaded/overloaded regimes) and user deployments (for diverse channel directions/strengths and CSIT qualities) \cite{Mao:2017}. \textit{In this work}, we build upon this last observation and show considering a simple two-user MISO BC with perfect CSIT that RS is a flexible framework for non-orthogonal transmission that generalizes, and subsumes as special cases, four seemingly completely different strategies, namely SDMA based on linear precoding, Orthogonal Multiple Access (OMA) where a resource is fully taken up by a single user, power-domain NOMA based on linearly precoded superposition coding with SIC, and physical-layer multicasting. This is the first paper to show analytically how RS unifies, outperforms, and specializes to SDMA, OMA, NOMA, and multicasting as a function of the disparity of the user channel strengths and the angle between the user channel directions. To that end, the paper differs from, and nicely complements, past works that analytically studied the rate performance of RS with imperfect CSIT \cite{Hao:2015,Dai:2016,Dai:2017} or looked at RS from an optimization perspective \cite{Mao:2017,Joudeh:2016a,Joudeh:2016b}. 

\emph{Notation:} $|.|$ and $\left\|.\right\|$ refer to the absolute value of a scalar and the $l_2$-norm of a vector. $\mathbf{I}$ is the identity matrix. $\mathbf{a}^{H}$ denotes the Hermitian transpose of vector $\mathbf{a}$. I.i.d. stands for independent and identically distributed. $\mathcal{CN}(0,\sigma^2)$ denotes the Circularly Symmetric Complex Gaussian distribution with zero mean and variance $\sigma^2$. $\sim$ stands for ``distributed as''. 
%\emph{Notations:} Italic letters and boldface lower-case letters denote scalars and vectors. $|.|$ and $\left\|.\right\|$ refer to the absolute value of a scalar and the 2-norm of a vector. The Hermitian transpose of vector $\mathbf{a}$ is denoted as $\mathbf{a}^{H}$. i.i.d. stands for independent and identically distributed. $\mathcal{CN}(0,\sigma^2)$ denotes Circularly Symmetric Complex Gaussian (CSCG) distribution. $\sim$ stands for ``distributed as''. 
\vspace{-0.5cm}
\section{System Model: Rate-Splitting Architecture}\label{system}
\par
\begin{figure}%[!t]
\centering
\includegraphics[width=0.9\columnwidth]{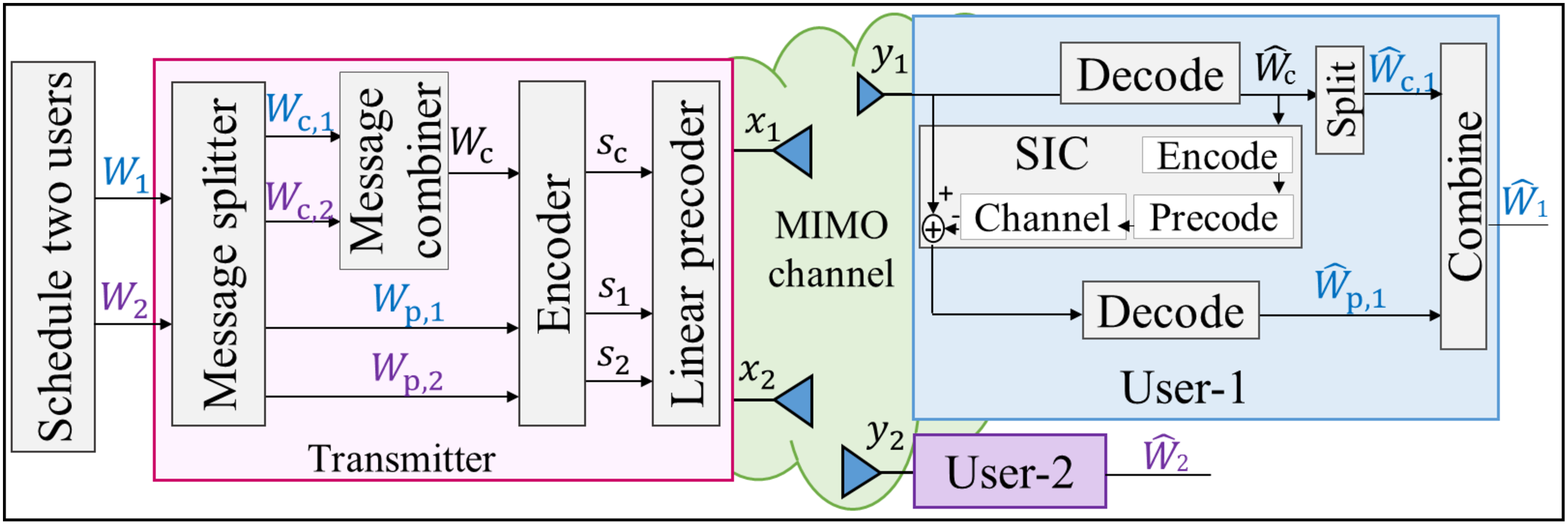}
\caption{Two-user system architecture with rate-splitting.}
\label{fig_sys_mod}
\vspace{-0.5cm}
\end{figure}

 We consider a MISO BC consisting of one transmitter with $n_t$ antennas and two single-antenna users. As per Fig. \ref{fig_sys_mod}, the architecture relies on rate-splitting of two messages $W_1$ and $W_2$ intended for user-1 and user-2, respectively. To that end, the message $W_k$ of user-$k$ is split into a common part $W_{\mathrm{c},k}$ and a private part $W_{\mathrm{p},k}$. The common parts $W_{\mathrm{c},1}, W_{\mathrm{c},2}$ of both users are combined into the common message $W_{\mathrm{c}}$, which is encoded into the common stream $s_{\mathrm{c}}$ using a codebook shared by both users. Hence, $s_{\mathrm{c}}$ is a common stream required to be decoded by both users, and contains parts of the messages $W_1$ and $W_2$ intended for user-1 and user-2, respectively. The private parts $W_{\mathrm{p},1}$ and $W_{\mathrm{p},2}$, respectively containing the remaining parts of the messages $W_1$ and $W_2$, are independently encoded into the private stream $s_1$ for user-1 and $s_2$ for user-2. Out of the two messages $W_1$ and $W_2$, three streams $s_{\mathrm{c}}$, $s_1$, and $s_2$ are therefore created. The streams are linearly precoded such that the transmit signal is given by
\begin{equation}
\mathbf{x}=\mathbf{p}_{\mathrm{c}} s_{\mathrm{c}}+\mathbf{p}_1 s_1+\mathbf{p}_2 s_2.
\end{equation}
Defining $\mathbf{s}=[s_{\mathrm{c}},s_1,s_2]^T$ and assuming that $\mathbb{E}[\mathbf{s}\mathbf{s}^H]=\mathbf{I}$, the average transmit power constraint is written as $P_{\mathrm{c}}+P_{1}+P_{2}\leq P$ where $P_{\mathrm{c}}=\left\|\mathbf{p}_{\mathrm{c}}\right\|^2$ and $P_{k}=\left\|\mathbf{p}_{k}\right\|^2$ with $k=1,2$. We refer to $\mathbf{h}_k$ as the channel vector of user-$k$, such that the signal received at user-$k$ can be written as
\begin{equation}
y_k=\mathbf{h}_k^H \mathbf{x}+n_k, \hspace{0.5cm}k=1,2,
\end{equation}
where $n_k\sim\mathcal{CN}(0,1)$ is Additive White Gaussian Noise (AWGN).
We further write the channel vectors as the product of their norm and direction as $\mathbf{h}_k=\left\|\mathbf{h}_k\right\| \bar{\mathbf{h}}_k$, and assume without loss of generality $\left\|\mathbf{h}_1\right\|\geq \left\|\mathbf{h}_2\right\|$. We also assume perfect CSI at the transmitter and the receivers.

\par At each user-$k$, the common stream $s_{\mathrm{c}}$ is first decoded into $\widehat{W}_{\mathrm{c}}$ by treating the interference from the private streams as noise. Using SIC, $\widehat{W}_{\mathrm{c}}$ is re-encoded, precoded, and subtracted from the received signal, such that user-$k$ can decode its private stream $s_k$ into $\widehat{W}_{\mathrm{p},k}$ by treating the remaining interference from the other private stream as noise. User-$k$ reconstructs the original message by extracting $\widehat{W}_{\mathrm{c},k}$ from $\widehat{W}_{\mathrm{c}}$, and combining $\widehat{W}_{\mathrm{c},k}$ with $\widehat{W}_{\mathrm{p},k}$ into $\widehat{W}_{k}$. Assuming Gaussian signalling and ideal SIC, the rate of the common stream is given by
\vspace{-0.2cm}
\begin{multline}
R_{\mathrm{c}}=\min\left(\log_2\left(1+\frac{\left|\mathbf{h}_1^H \mathbf{p}_{\mathrm{c}}\right|^2}{1+\left|\mathbf{h}_1^H \mathbf{p}_{1}\right|^2+\left|\mathbf{h}_1^H \mathbf{p}_{2}\right|^2}\right),\right.\\
\left.\log_2\left(1+\frac{\left|\mathbf{h}_2^H \mathbf{p}_{\mathrm{c}}\right|^2}{1+\left|\mathbf{h}_2^H \mathbf{p}_{1}\right|^2+\left|\mathbf{h}_2^H \mathbf{p}_{2}\right|^2}\right)\right),
\end{multline}
and the rates of the two private streams are obtained as
%\begin{equation}
%\begin{aligned}
%R_{1}&=\log_2\left(1+\frac{\left|\mathbf{h}_1^H \mathbf{p}_{1}\right|^2}{1+\left|\mathbf{h}_1^H \mathbf{p}_{2}\right|^2}\right),\\
%R_{2}&=\log_2\left(1+\frac{\left|\mathbf{h}_2^H \mathbf{p}_{2}\right|^2}{1+\left|\mathbf{h}_2^H \mathbf{p}_{1}\right|^2}\right).
%\end{aligned}
%\end{equation}
\begin{equation}
R_{k}=\log_2\left(1+\frac{\left|\mathbf{h}_k^H \mathbf{p}_{k}\right|^2}{1+\left|\mathbf{h}_k^H \mathbf{p}_{j}\right|^2}\right), k \neq j.
\end{equation}
The rate of user-$k$ is given by $R_k+R_{\mathrm{c},k}$ where $R_{\mathrm{c},k}$ is the rate of the common part of the $k$th user’s message, i.e., $W_{\mathrm{c},k}$, and it satisfies $R_{\mathrm{c},1}+R_{\mathrm{c},2}=R_{\mathrm{c}}$. The sum-rate is therefore simply written as $R_{\mathrm{s}}=\sum_{k=1,2} R_k+R_{\mathrm{c},k}=R_{\mathrm{c}}+R_{1}+R_{2}$.

\par
\begin{figure}%[!t]
\centering
\includegraphics[width=0.9\columnwidth]{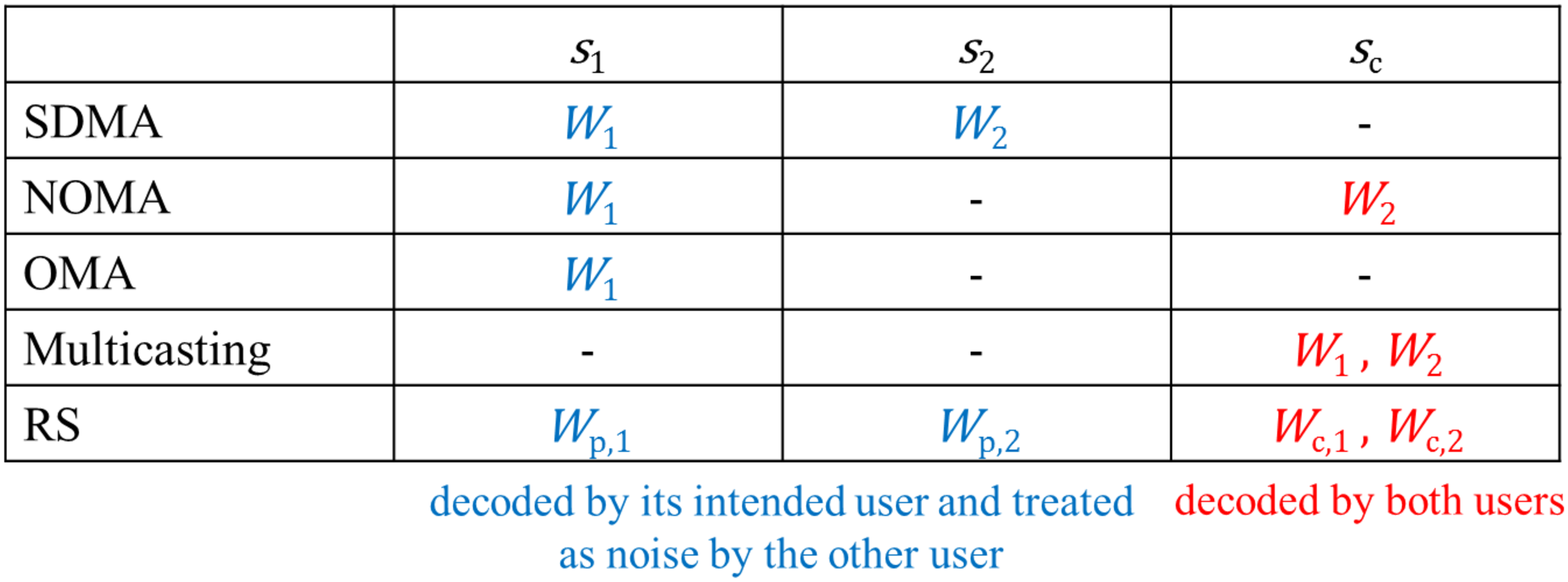}
\vspace{-0.4cm}
\caption{Mapping of messages to streams.}
\label{fig_mapping}
\vspace{-0.5cm}
\end{figure}

 By adjusting the message split and the power allocation to the common stream and the private streams, RS enables the decoding of part of the interference (thanks to the presence of the common stream) and treating the remaining part (the private stream of the other user) as noise. Therefore, the introduced RS architecture allows the exploration of a wide range of strategies. Among those strategies, there are four extreme cases, namely, SDMA, NOMA, OMA, and physical-layer multicasting. Indeed, SDMA is obtained by allocating no power to the common stream ($P_{\mathrm{c}}=0$) such that $W_k$ is encoded directly into $s_k$. No interference is decoded at the receiver using the common message, and the interference between $s_1$ and $s_2$ is fully treated as noise. NOMA is obtained by encoding $W_2$ entirely into $s_{\mathrm{c}}$ (i.e., $W_{\mathrm{c}}=W_2$) and $W_1$ into $s_{1}$, and turning off $s_2$ ($P_{2}=0$). In this way, user-1 fully decodes the interference created by the message of user-2. OMA is a sub-strategy of SDMA and NOMA and is obtained when only user-1 (with the stronger channel gain) is scheduled ($P_{\mathrm{c}}=0, P_2=0$). Multicasting is obtained by combining and encoding both $W_1$ and $W_2$ into $s_{\mathrm{c}}$, and turning off $s_1$ and $s_2$ ($P_1=0, P_2=0$). The mapping of the messages to the streams is further illustrated in Fig. \ref{fig_mapping}.

\begin{remark} Recall that the maximum number of interference-free streams (also called Degrees-of-Freedom DoF) in a two-user MISO BC is equal to 2. From the above system model, both SDMA and RS can achieve such a DoF by precoding $s_1$ and $s_2$ using zero-forcing (ZF). On the other hand, OMA, NOMA, and multicasting can achieve at most a DoF of 1 (irrespectively of how the precoders and power allocation are optimized), which leads to a rate loss at high Signal-to-Noise Ratio (SNR) in general multi-antenna settings, as already highlighted in \cite{Joudeh:2017,Mao:2017}. 
\end{remark}

\vspace{-0.3cm}
\section{Sum-Rate Analysis}\label{precoder_section}
Our objective is to derive tractable and insightful sum-rate expressions to illustrate the flexibility of RS in unifying SDMA, OMA, NOMA, and multicasting. To that end, we do not optimize the precoding directions jointly with the power allocation as in \cite{Joudeh:2016a,Mao:2017} but rather fix the precoding directions using ZF for the private streams, and adjust the power allocation among all the streams\footnote{Simulations in Section \ref{evaluations} show that the conclusions drawn with the simple precoders also hold with the numerically optimized precoders of \cite{Joudeh:2016a,Mao:2017}.}. This leads to $\left|\mathbf{h}_2^H \mathbf{p}_{1}\right|=0$, $\left|\mathbf{h}_1^H \mathbf{p}_{2}\right|=0$, and $\left|\mathbf{h}_k^H \mathbf{p}_{k}\right|^2=\left\|\mathbf{h}_k\right\|^2 \rho P_k$, $k=1,2$, where $\rho=1-\left|\bar{\mathbf{h}}_1^H\bar{\mathbf{h}}_2\right|^2$ ($\rho=0$ corresponds to aligned channels and $\rho=1$ to orthogonal channels). The precoder of the common stream is then to be designed such that 
\begin{equation}
\max_{\mathbf{p}_{\mathrm{c}}} \min\left(\frac{\left|\mathbf{h}_1^H \mathbf{p}_{\mathrm{c}}\right|^2}{1+\left|\mathbf{h}_1^H \mathbf{p}_{1}\right|^2},\frac{\left|\mathbf{h}_2^H \mathbf{p}_{\mathrm{c}}\right|^2}{1+\left|\mathbf{h}_2^H \mathbf{p}_{2}\right|^2}\right). 
\end{equation}
Defining $\gamma_k^2=1+\left|\mathbf{h}_k^H \mathbf{p}_{k}\right|^2=1+\left\|\mathbf{h}_k\right\|^2 \rho P_k$, $k=1,2$, and $\tilde{\mathbf{h}}_k=\mathbf{h}_k/\gamma_k$, the problem is re-written as 
\begin{equation}\label{common_precoder_design}
\max_{\mathbf{p}_{\mathrm{c}}} \min\left(\big|\tilde{\mathbf{h}}_1^H \mathbf{p}_{\mathrm{c}}\big|^2,\big|\tilde{\mathbf{h}}_2^H \mathbf{p}_{\mathrm{c}}\big|^2\right). 
\end{equation}
Following \cite{Hsiao:2015}, the solution of \eqref{common_precoder_design} is $\mathbf{p}_{\mathrm{c}}=\sqrt{P_{\mathrm{c}}}\mathbf{f}_{\mathrm{c}}$ with the precoder direction $\mathbf{f}_{\mathrm{c}}$ ($\left\|\mathbf{f}_{\mathrm{c}}\right\|^2=1$) given by
\begin{equation}
\mathbf{f}_{\mathrm{c}}=\frac{1}{\sqrt{\lambda}}\left(\mu_1 \tilde{\mathbf{h}}_1 + \mu_2 \tilde{\mathbf{h}}_2 e^{-j \angle\alpha_{12}}\right),\label{opt_pc}
\end{equation}
where
\begin{equation}
\lambda =\frac{\alpha_{11} \alpha_{22}-\left|\alpha_{12}\right|^2}{\alpha_{11} + \alpha_{22}- 2 \left|\alpha_{12}\right|}, 
\end{equation}
\begin{equation}
\left[\begin{array}{c}\mu_1 \\ \mu_2\end{array}\right] =\frac{1}{\alpha_{11} + \alpha_{22}- 2 \left|\alpha_{12}\right|}\left[\begin{array}{c} \alpha_{22}- \left|\alpha_{12}\right| \\ \alpha_{11}- \left|\alpha_{12}\right|\end{array}\right],
\end{equation}
\begin{equation}
\left[\begin{array}{cc}\alpha_{11} & \alpha_{12} \\ \alpha_{12}^* & \alpha_{22}\end{array}\right]=\left[\begin{array}{c} \tilde{\mathbf{h}}_1^H \\ \tilde{\mathbf{h}}_2^H \end{array}\right]\left[\begin{array}{cc}\tilde{\mathbf{h}}_1 & \tilde{\mathbf{h}}_2\end{array}\right].
\end{equation}
\vspace{-0.5cm}
\subsection{Sum-Rate at Finite SNR}
The sum-rate with the above precoder designs can be written as $R_{\mathrm{s}}=R_{\mathrm{c}}+\log_2\left(\gamma_1^2\right)+\log_2\left(\gamma_2^2\right)$,
where $R_{\mathrm{c}}\!=\!\min\!\big(\log_2\big(1\!+\!\big|\tilde{\mathbf{h}}_1^H \mathbf{p}_{\mathrm{c}}\big|^2\big),\log_2\big(1\!+\!\big|\tilde{\mathbf{h}}_2^H \mathbf{p}_{\mathrm{c}}\big|^2\big)\big)$. With $\mathbf{p}_{\mathrm{c}}$ as per \eqref{opt_pc}, following \cite{Hsiao:2015}, $\big|\tilde{\mathbf{h}}_1^H \mathbf{p}_{\mathrm{c}}\big|\!=\!\big|\tilde{\mathbf{h}}_2^H \mathbf{p}_{\mathrm{c}}\big|$, and we can write $R_{\mathrm{c}}=\log_2\big(1+\big|\tilde{\mathbf{h}}_2^H \mathbf{p}_{\mathrm{c}}\big|^2\big)$, and the sum-rate simply as
\begin{equation}
R_{\mathrm{s}}=\log_2\left(\gamma_1^2\right)+\log_2\left(\gamma_2^2+\big|\mathbf{h}_2^H \mathbf{p}_{\mathrm{c}}\big|^2\right).\label{sumrate}
\end{equation}
%&=\log_2\left(1+\frac{\big|\mathbf{h}_2^H \mathbf{p}_{\mathrm{c}}\big|^2}{\gamma_2^2}\right)+\log_2\left(\gamma_1^2\right)+\log_2\left(\gamma_2^2\right),\nonumber\\
\par Consider a fraction $t$ of the total transmit power $P$ is allocated to the private streams such that $P_1+P_2=tP$ and the remaining power $P_{\mathrm{c}}=\left(1-t\right)P$ is allocated to the common stream. For a given $t$, the optimal values of $P_1$ and $P_2$, maximizing the sum-rate of the private streams, are given by the Water-Filling (WF) solution
\begin{equation}
P_k=\max\left(\mu-\frac{1}{\left\|\mathbf{h}_k\right\|^2 \rho},0\right), \hspace{0.3cm} k=1,2, \label{WF_solution}
\end{equation}
with the water level $\mu$ chosen such that $P_1\!+\!P_2\!=\!tP$, and set as $\mu\!=\!\frac{tP}{2}\!+\!\frac{1}{2\rho}\left[\frac{1}{\left\|\mathbf{h}_1\right\|^2}\!+\!\frac{1}{\left\|\mathbf{h}_2\right\|^2}\right]$ in the sequel. Let us also introduce 
$\Gamma\!=\!\frac{1}{\rho}\left[\frac{1}{\left\|\mathbf{h}_2\right\|^2}\!-\!\frac{1}{\left\|\mathbf{h}_1\right\|^2}\right]$,
which is a function of two main parameters: $\rho$ reflecting the angle between the user channel directions, and $\frac{1}{\left\|\mathbf{h}_2\right\|^2}-\frac{1}{\left\|\mathbf{h}_1\right\|^2}$ reflecting the disparity of the channel strengths.
We can then identify two main regimes.

\subsubsection{OMA/NOMA/Multicasting Regime}
%\begin{enumerate}
%\item \textit{OMA/NOMA/Multicasting Regime}: 
If $\mu \leq \frac{1}{\left\|\mathbf{h}_2\right\|^2 \rho}$, i.e., $tP \leq \Gamma$, we set $P_2=0$ and $P_1=tP$ according to \eqref{WF_solution}, and RS specializes to multicasting for $t=0$, NOMA for $0<t<1$, and OMA for $t=1$. In this regime, $t$ needs to be adjusted so as to identify the best strategy among OMA, NOMA, and multicasting, and therefore efficiently allocate power across the common stream $s_{\mathrm{c}}$ and the private stream $s_1$. 
\subsubsection{RS/SDMA Regime}
%\item \textit{RS/SDMA Regime}: 
If $\mu > \frac{1}{\left\|\mathbf{h}_2\right\|^2 \rho}$, i.e. $tP > \Gamma$, the WF solution \eqref{WF_solution} leads to $P_1=\mu-\frac{1}{\left \|\mathbf{h}_1\right \|^2\rho}=\frac{tP}{2}+\frac{\Gamma }{2}>0$ and $P_2=\mu-\frac{1}{\left \|\mathbf{h}_2\right \|^2\rho}=\frac{tP}{2}-\frac{\Gamma }{2}>0$. RS specializes to SDMA whenever $t$ is set to 1, but does not specializes to any other known scheme for $0<t<1$. In this regime, $t$ needs to be adjusted, as explained in the sequel, so as to allocate the power efficiently across the common stream and the two private streams. 
Substituting the expressions of $P_k$ and $\gamma_k^2$, $k=1,2$, into \eqref{sumrate}, we can write
\begin{equation}
R_{\mathrm{s}}=\log_2\left(ac+\left(ad+bc\right)t+bdt^2\right),
\end{equation} 
where $b=\frac{\left \|\mathbf{h}_1\right \|^2\rho P}{2}$, $a=1+\frac{\Gamma}{P}b$, $d=\frac{\left \|\mathbf{h}_2\right \|^2\rho P}{2}-|\mathbf{h}_2^H\mathbf{f}_c|^2P$, and $c=1-\frac{\Gamma}{P}d+|\mathbf{h}_2^H\mathbf{f}_c|^2(P-\Gamma)$. The value of $t$ that maximizes $R_{\mathrm{s}}$ is the solution of $\frac{\partial R_s}{\partial t}=0$, which is written as $t=-\frac{a}{2b}-\frac{c}{2d}$. Since $t\leq 1$, the optimal value $t^{\star}$ is given in closed form by \eqref{eq: t opt} at the top of the next page. For $t^{\star}<1$, RS yields a non-zero sum-rate enhancement over SDMA.
\begin{table*}
\begin{equation}\label{eq: t opt}
t^{\star}=\min\left(-\frac{a}{2b}-\frac{c}{2d},1\right)=\min\left(\frac{\left|\bar{\mathbf{h}}_2^H\mathbf{f}_c\right|^2}{2\left|\bar{\mathbf{h}}_2^H\mathbf{f}_c\right|^2-\rho}+\frac{1}{2\rho}\left(\frac{1}{\left \|\mathbf{h}_1\right \|^2}+\frac{1}{\left \|\mathbf{h}_2\right \|^2}\right)\left(\frac{{2\rho-2{\left|\bar{\mathbf{h}}_2^H\mathbf{f}_c\right|^2}}}{ 2{\left|\bar{\mathbf{h}}_2^H\mathbf{f}_c\right|^2}-{\rho}}\right)\frac{1}{P},1\right).
\end{equation}
\hrulefill
\vspace{-5mm}
\end{table*}

\begin{remark} It is important to note that the solution $t=-\frac{a}{2b}-\frac{c}{2d}$ holds because the coefficients $a$, $b$, $c$, $d$ are not functions of $t$. This could appear surprising since $c$ and $d$ are functions of $\mathbf{f}_c$, which, according to \eqref{common_precoder_design}, is a function of $P_1$ and $P_2$ and therefore of $t$. However, interestingly, in the regime where $P_1>0$ and $P_2>0$, we can show that $\mathbf{f}_c$ is not a function of $t$. Making use of $P_1=\frac{tP}{2}+\frac{\Gamma }{2}$ and $P_2=\frac{tP}{2}-\frac{\Gamma }{2}$, we can write $\gamma_k^2=1+\left \|\mathbf{h}_k\right \|^2\rho P_k=\frac{f(t)}{\left \|\mathbf{h}_j\right \|^2}$, $k,j=1,2$ and $k \neq j$, with $f(t)=\frac{\left \|\mathbf{h}_1\right \|^2+\left \|\mathbf{h}_2\right \|^2+{\left \|\mathbf{h}_1\right \|^2\left \|\mathbf{h}_2\right \|^2\rho P}t}{2}$. We then obtain
\begin{equation}
\begin{aligned}
	&\max_{\mathbf{f}_{\mathrm{c}}} \min\left(\big|\tilde{\mathbf{h}}_1^H \mathbf{f}_{\mathrm{c}}\big|^2,\big|\tilde{\mathbf{h}}_2^H \mathbf{f}_{\mathrm{c}}\big|^2\right)\\
	\Leftrightarrow &  \max_{\mathbf{f}_c}  \min \left(\gamma_2^2{\big|\mathbf{h}_1^H\mathbf{f}_c\big|^2},{\gamma_1^2\big|\mathbf{h}_2^H\mathbf{f}_c\big|^2}\right)\\
	\Leftrightarrow&
\max_{\mathbf{f}_c}  \min \left(f(t){\big|\bar{\mathbf{h}}_1^H\mathbf{f}_c\big|^2},{f(t)\big|\bar{\mathbf{h}}_2^H\mathbf{f}_c\big|^2}\right)\\
\Leftrightarrow&
\max_{\mathbf{f}_c}  \min \left({\big|\bar{\mathbf{h}}_1^H\mathbf{f}_c\big|^2},{\big|\bar{\mathbf{h}}_2^H\mathbf{f}_c\big|^2}\right),\label{fc_simple_design}
\end{aligned}
\end{equation}
which reveals that $\mathbf{f}_{\mathrm{c}}$ is not a function of $t$ and the channel strength disparity, but only of the channel directions.
\end{remark}

%\end{enumerate}
\vspace{-0.3cm}
\subsection{Sum-Rate at High SNR}\label{high_snr_analysis}
At high SNR, considering $0 < t \leq 1$ and $\rho>0$, the solution in \eqref{WF_solution} allocates power uniformly across the two private streams as $P_1=P_2=\frac{tP}{2}>0$. Hence, only RS and SDMA are suitable strategies at high SNR. 
%This uniform allocation also leads at high SNR to a simpler design of $\mathbf{f}_{\mathrm{c}}$ in \eqref{common_precoder_design} and \eqref{opt_pc} as
%\begin{equation}
%\max_{\mathbf{f}_{\mathrm{c}}} \min\left(\big|\bar{\mathbf{h}}_1^H \mathbf{f}_{\mathrm{c}}\big|^2,\big|\bar{\mathbf{h}}_2^H \mathbf{f}_{\mathrm{c}}\big|^2 \right),
%\end{equation}
%which highlights that $\mathbf{f}_{\mathrm{c}}$ is not a function of $t$ and the channel strength disparity in such regime.
The sum-rate in \eqref{sumrate} can then be written as
\begin{equation}
R_{\mathrm{s}}\stackrel{P\nearrow}{=}\log_2\big(\left\|\mathbf{h}_1\right\|^2 \! \rho \big)+2\log_2\left(P\right)+\log_2\left(e t^2+f t\right)\label{RS_highSNR}
\end{equation}
%&\stackrel{P\nearrow}{=}\log_2\left(\left\|\mathbf{h}_1\right\|^2 \! \rho P_1\right)+\log_2\left(\left\|\mathbf{h}_2\right\|^2 \rho P_2+\big|\mathbf{h}_2^H \mathbf{f}_{\mathrm{c}}\big|^2P_{\mathrm{c}}\right),\nonumber\\
with $e=\frac{\left\|\mathbf{h}_2\right\|^2 \rho}{4}\!-\!\frac{\left|\mathbf{h}_2^H \mathbf{f}_{\mathrm{c}}\right|^2}{2}$, $f=\frac{\left|\mathbf{h}_2^H \mathbf{f}_{\mathrm{c}}\right|^2}{2}$.
Not surprisingly, a DoF of 2 is achieved in \eqref{RS_highSNR}. More interesting is the fact that RS brings a constant sum-rate enhancement over SDMA. Indeed, the value of $t$ that maximizes \eqref{RS_highSNR} is given by
\begin{equation}\label{optt_highsnr}
t^{\star}=\min\left(\frac{-f}{2e},1\right)=\min\left(\frac{\big|\bar{\mathbf{h}}_2^H \mathbf{f}_{\mathrm{c}}\big|^2}{2\big|\bar{\mathbf{h}}_2^H \mathbf{f}_{\mathrm{c}}\big|^2-\rho},1\right),
\end{equation}
%where operator $\left(x\right)^1=\left\{\begin{array}{l}
%x, \hspace{0.2cm} x\leq 1 \\
%1, \hspace{0.2cm} x>1 
%\end{array}
%\right.$, 
which coincides with \eqref{eq: t opt} when $P\rightarrow \infty$, and leads to a high SNR non-zero (whenever $0<t^{\star}<1$) sum-rate gap between RS and SDMA ($t=1$) given by
%\begin{align}
%\Delta R_{\mathrm{s}}&=\left.R_{\mathrm{s}}\right|_{t^{\star}}-\left.R_{\mathrm{s}}\right|_{t=1}\\
%&=\log_2\left(\frac{\big|\bar{\mathbf{h}}_2^H \mathbf{f}_{\mathrm{c}}\big|^4}{\rho\left(2\big|\bar{\mathbf{h}}_2^H \mathbf{f}_{\mathrm{c}}\big|^2-\rho\right)}\right),\hspace{0.3cm} \textnormal{if}\hspace{0.2cm} 0<t^{\star}<1. \label{delta_Rs}
%\end{align}
\begin{equation}
\Delta R_{\mathrm{s}}=\left.R_{\mathrm{s}}\right|_{t^{\star}}-\left.R_{\mathrm{s}}\right|_{t=1}=\log_2\left(\frac{\big|\bar{\mathbf{h}}_2^H \mathbf{f}_{\mathrm{c}}\big|^4}{\rho\left(2\big|\bar{\mathbf{h}}_2^H \mathbf{f}_{\mathrm{c}}\big|^2-\rho\right)}\right). \label{delta_Rs}
\end{equation}
$t^{\star}$ increases and $\Delta R_{\mathrm{s}}$ decreases as $\rho$ increases, and both are not a function of the channel strengths. The sum-rate gap between RS and NOMA/OMA/multicasting grows unbounded as $P\!\rightarrow\!\infty$ due to the difference in DoF (Remark 1).

\vspace{-0.4cm}
\subsection{Discussions}\label{discussion}
\par We can draw several insights from the above analysis. \textit{First}, for given $t$, $\rho$, $\left\|\mathbf{h}_1\right\|^2$, and $\left\|\mathbf{h}_2\right\|^2$, as $P$ increases, the SNRs of the private streams increase, while the Signal-to-Interference-plus-Noise Ratio (SINR) of the common stream ultimately saturates (interference limited regime). This suggests that the common message can only provide a constant rate improvement at high SNR, while the two private streams provide the DoF of 2. \textit{Second}, the quantity $\rho$ is present in the SNRs of both private streams and has the effect of increasing/decreasing the SNRs of those two streams. A lower $\rho$ indicates that both private streams effectively operate at a lower SNR. According to \eqref{WF_solution}, for a given $t$, a low $\rho$ favors power allocation to a single private stream (NOMA/OMA/Multicasting regime) over a wider range of $P$, and also leads to a smaller interference power (and therefore a higher rate) for the common stream. A higher $\rho$ leads to a higher effective SNR and therefore a better capability to support two private streams (RS/SDMA regime). \textit{Third}, as the disparity of channel strengths increases, the WF solution allocates a larger amount of power to the stronger user (user-1) over a wider range of $P$ (for a given $t$). Beyond a certain disparity, for given $t$, $P$, and $\rho$, $P_2$ is turned off and RS specializes to NOMA/OMA.
\vspace{-0.4cm}
\section{Evaluations}\label{evaluations}

\begin{figure}%[!t]
%\centering
\begin{subfigmatrix}{2}
\subfigure[Optimum $t$]{\label{a}\includegraphics[width = 4.3cm]{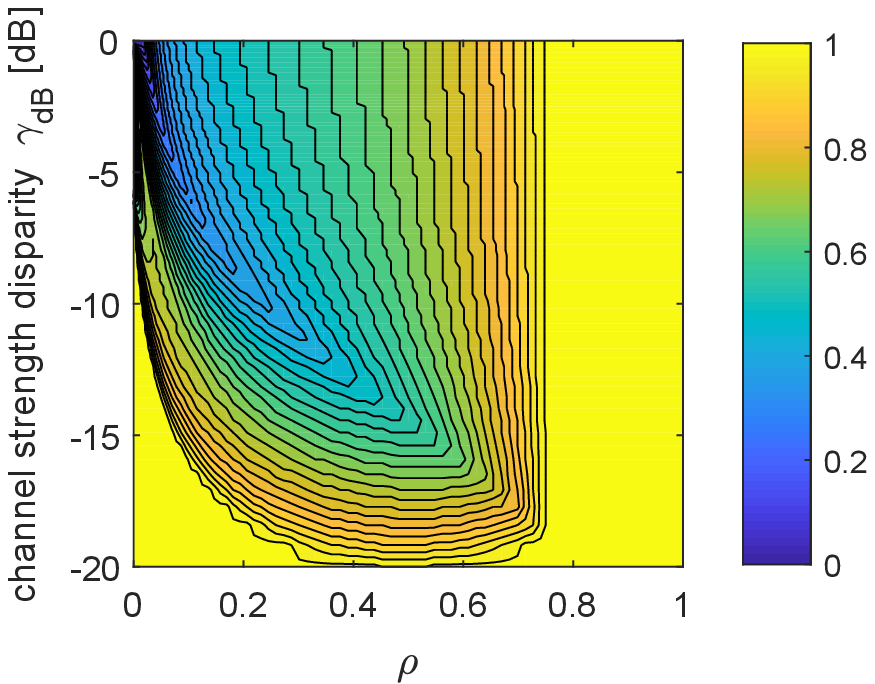}}
\subfigure[Regions of operation]{\label{b}\includegraphics[width = 4.3cm]{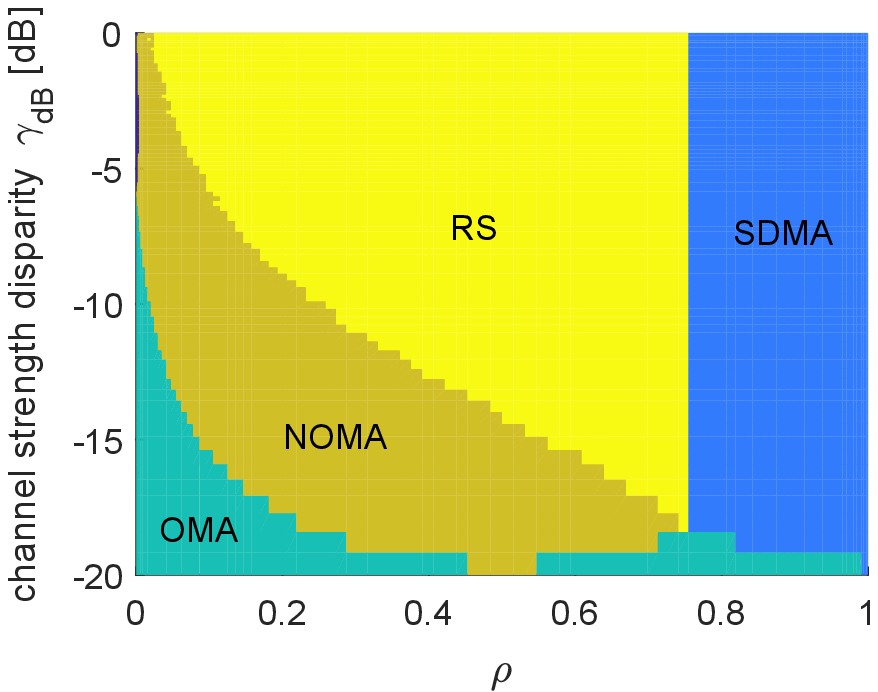}}
\end{subfigmatrix}
\caption{Optimum $t$ in (a) and regions of operation for RS, SDMA, NOMA, and OMA in (b). Precoding strategies from Section \ref{precoder_section} with $P=100$W.}
\label{topt}
\vspace{-0.3cm}
\end{figure}

%\begin{figure}%[!t]
%%\centering
%\begin{subfigmatrix}{2}
%\subfigure[Multicast and ZF precoding]{\label{a}\includegraphics[width = 4.3cm]{Fig3Thetapi100Gamma001_ZF_test.eps}}
%\subfigure[WMMSE Opt. $\epsilon=10^{-3}$]{\label{b}\includegraphics[width = 4.3cm]{Fig3Thetapi100Gamma001_WMMSEaccuracy0001_test.eps}}
%\end{subfigmatrix}
%\caption{Optimum $t$ for precoding strategies of Section \ref{precoder_section} in (a) and based on WMMSE optimization in (b). $P=100$W.}
%\label{topt}
%\end{figure}
%
%\begin{figure}%[!t]
%%\centering
%\begin{subfigmatrix}{2}
%\subfigure[Multicast and ZF precoding]{\label{a}\includegraphics[width = 4.3cm]{Fig4Thetapi100Gamma001_ZF_test.eps}}
%\subfigure[WMMSE Opt. $\epsilon=10^{-3}$]{\label{b}\includegraphics[width = 4.3cm]{Fig4Thetapi100Gamma001_WMMSEaccuracy0001_test.eps}}
%\end{subfigmatrix}
%\caption{Regions of operation for RS, SDMA, NOMA, and OMA with precoders from Section \ref{precoder_section} (a) and based on WMMSE opt. (b). $P=100$W.}
%\label{regions}
%\end{figure}

\begin{figure}%[!t]
%\centering
\begin{subfigmatrix}{2}
\subfigure[$P=10$ W (SNR=10dB)]{\label{a}\includegraphics[width = 4.3cm]{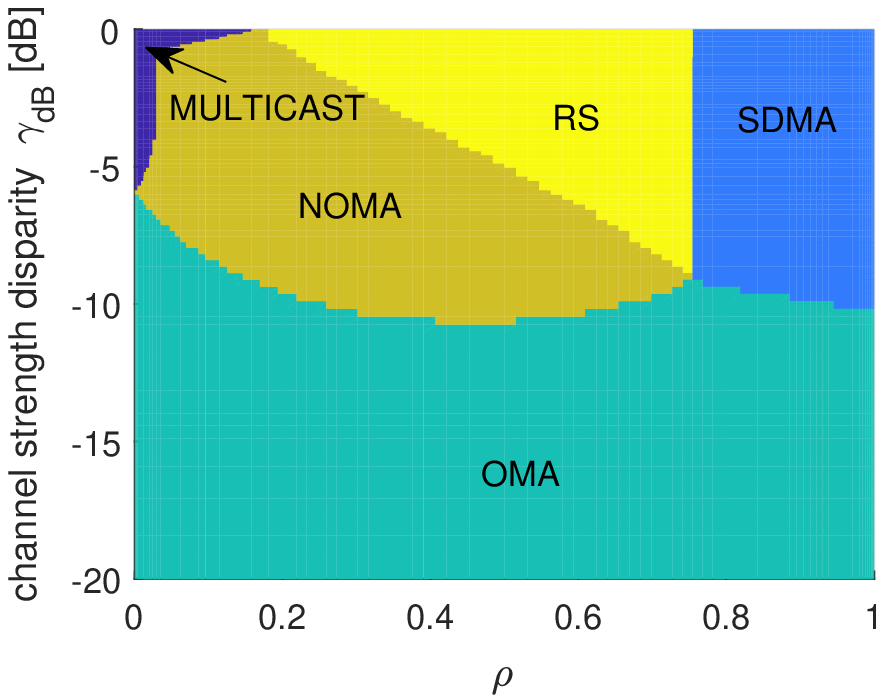}}
\subfigure[$P=1000$ W (SNR=30dB)]{\label{b}\includegraphics[width = 4.3cm]{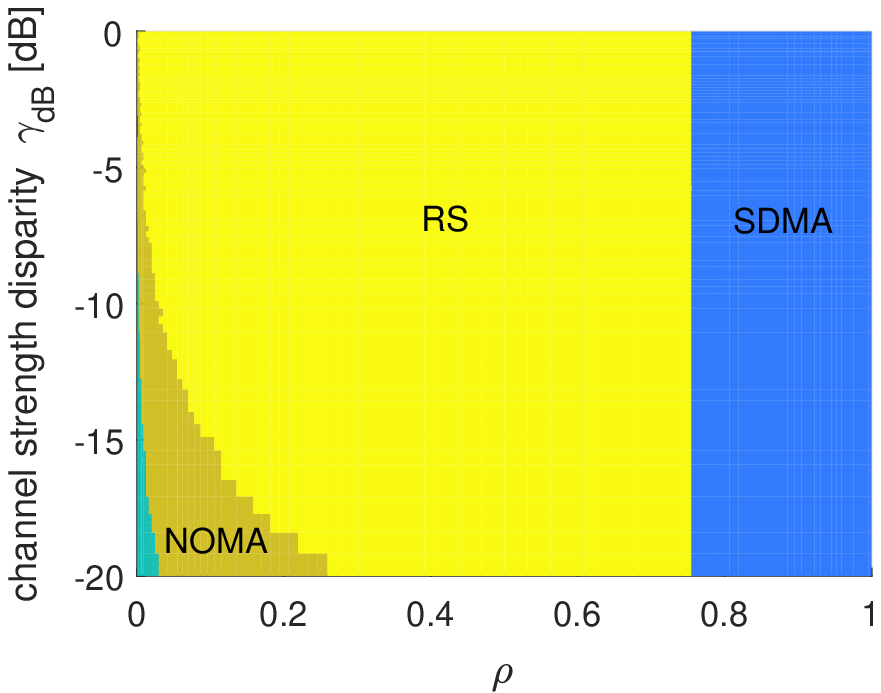}}
\end{subfigmatrix}
\caption{Regions of operation for RS, SDMA, NOMA, OMA and Multicast with precoders from Section \ref{precoder_section} for $P=10 \textnormal{W}, 1000 \textnormal{W}$.}
\label{regions_P}
\vspace{-0.4cm}
\end{figure}

\begin{figure*}%[!t]
%\centering
\begin{subfigmatrix}{3}
\subfigure[$P=10$ W (SNR=10dB)]{\label{a}\includegraphics[width = 5.5cm]{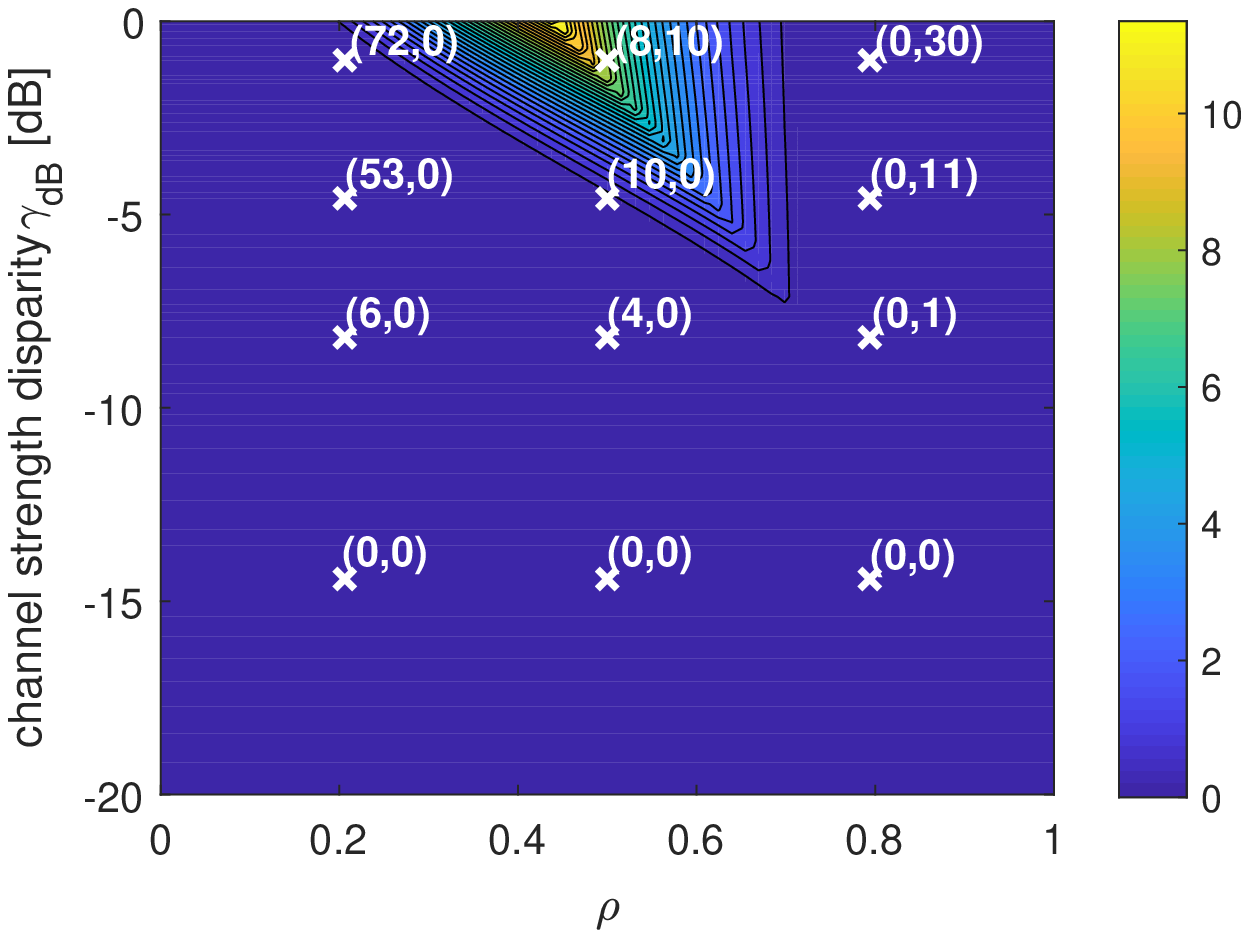}}
\subfigure[$P=100$ W (SNR=20dB)]{\label{b}\includegraphics[width = 5.5cm]{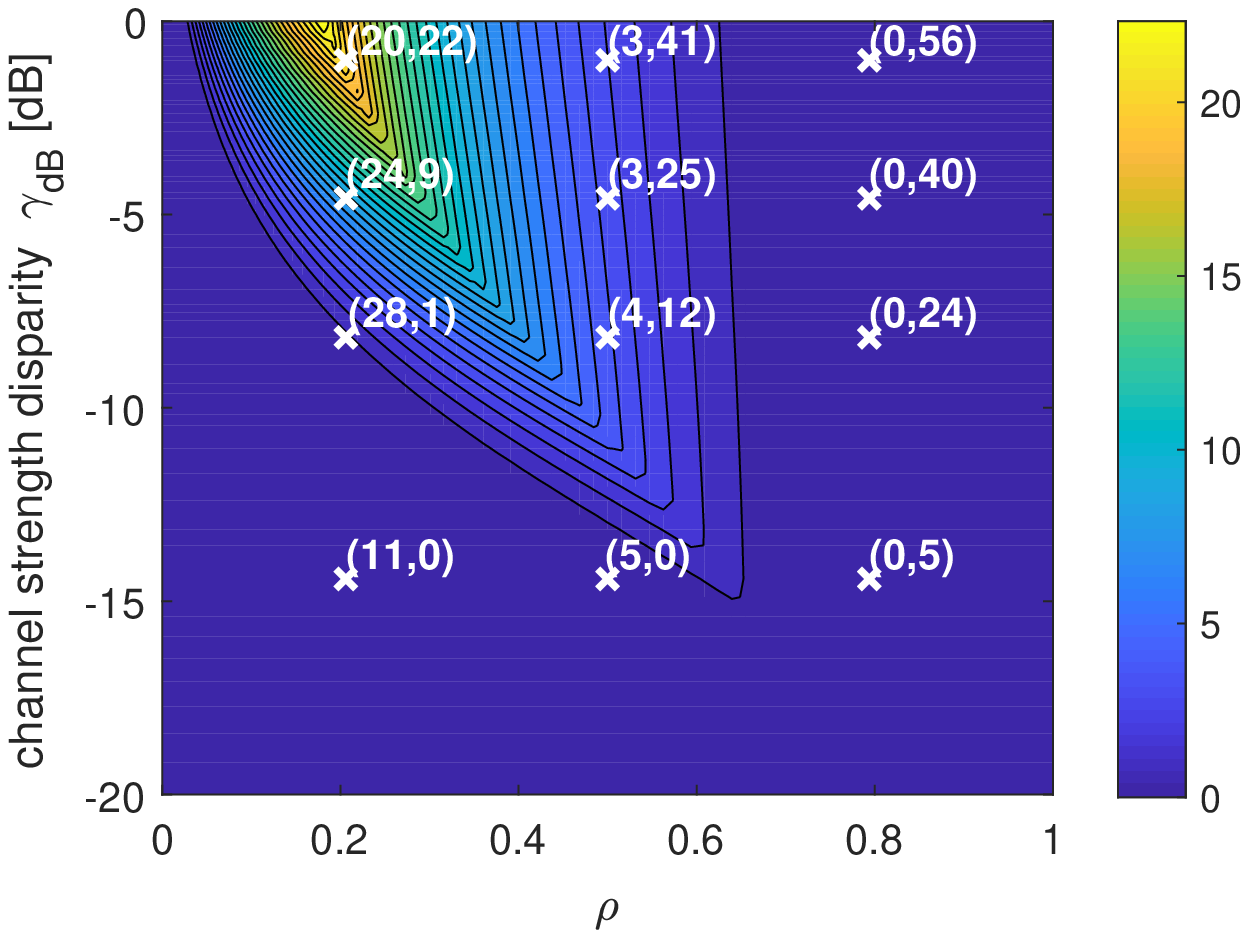}}
\subfigure[$P=1000$ W (SNR=30dB)]{\label{c}\includegraphics[width = 5.5cm]{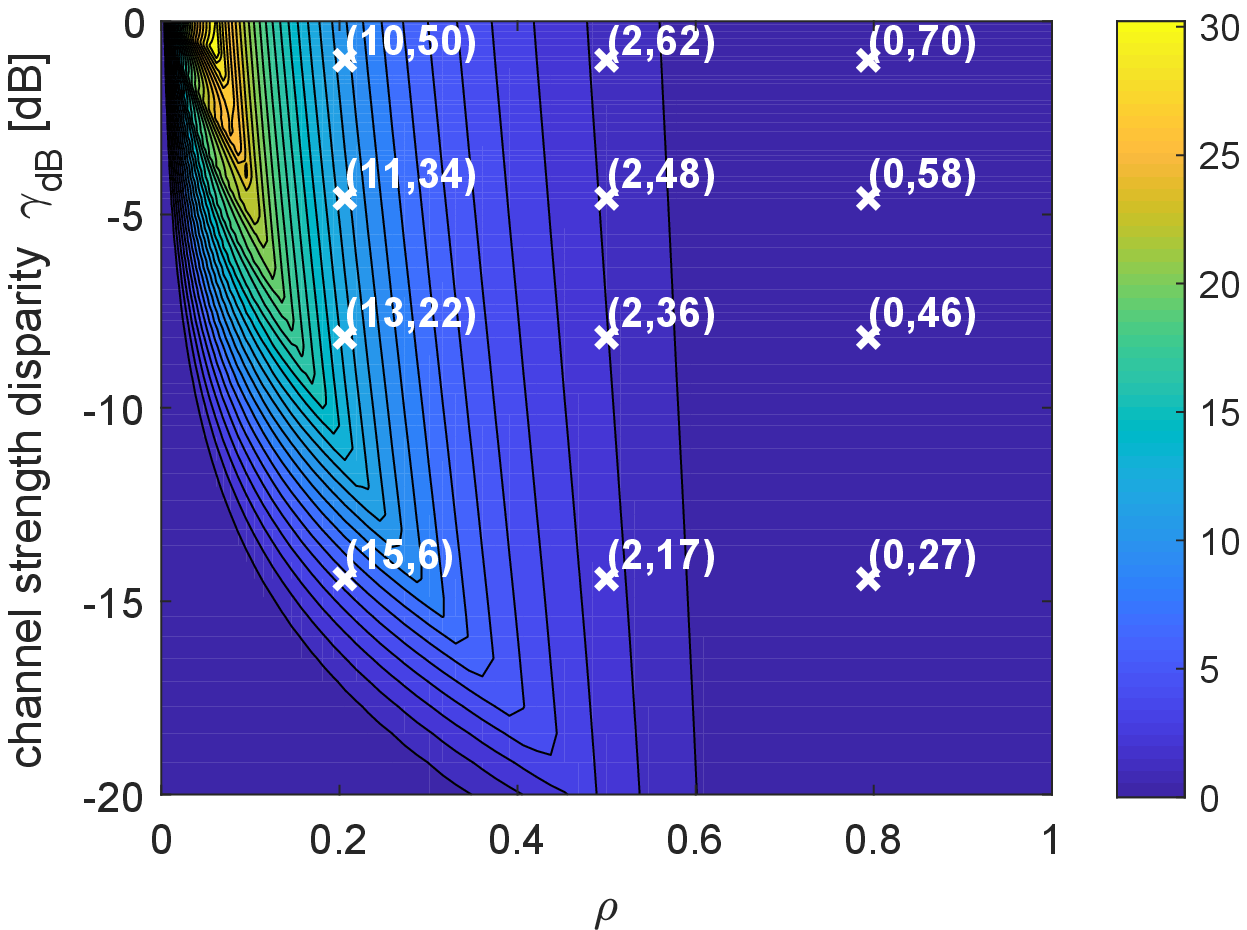}}
\end{subfigmatrix}
\vspace{-0.2cm}
\caption{Relative sum-rate gain [\%] of RS over dynamic switching between SDMA and NOMA, with $n_t=2$ and precoders from Section \ref{precoder_section}. The values in brackets indicate sum-rate gains over SDMA and NOMA, respectively.}
\label{gain}
\vspace{-0.3cm}
\end{figure*}

\begin{figure*}%[!t]
%\centering
\begin{subfigmatrix}{3}
\subfigure[$u_1=10^{0.5}, u_2=1$]{\label{a}\includegraphics[width = 5.5cm]{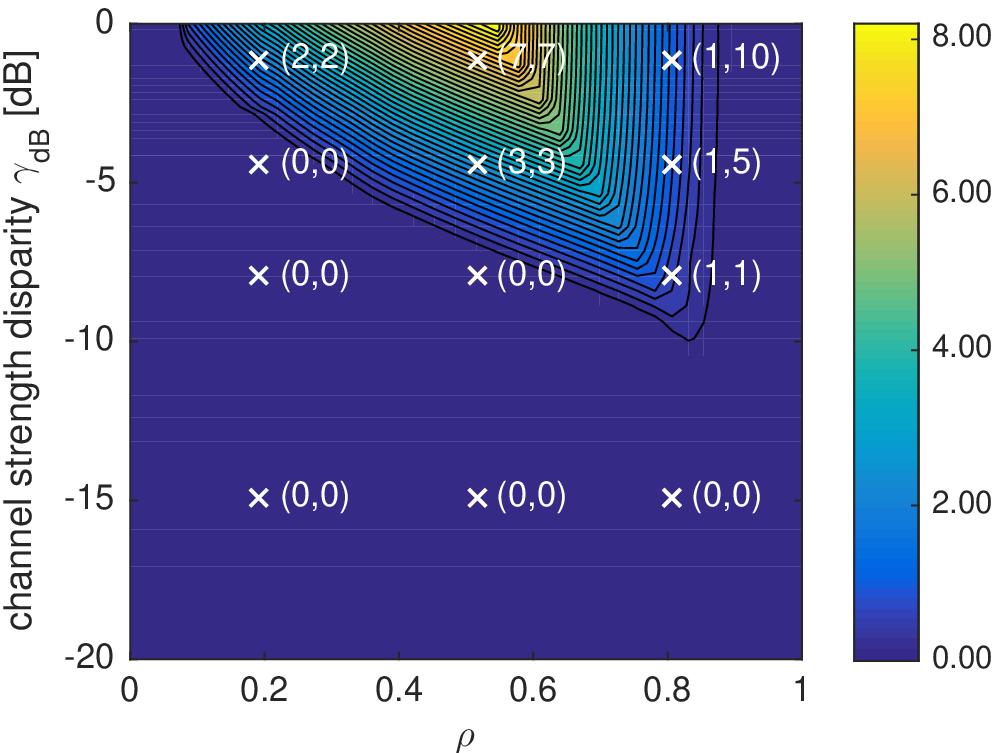}}
\subfigure[$u_1=1, u_2=1$]{\label{b}\includegraphics[width = 5.5cm]{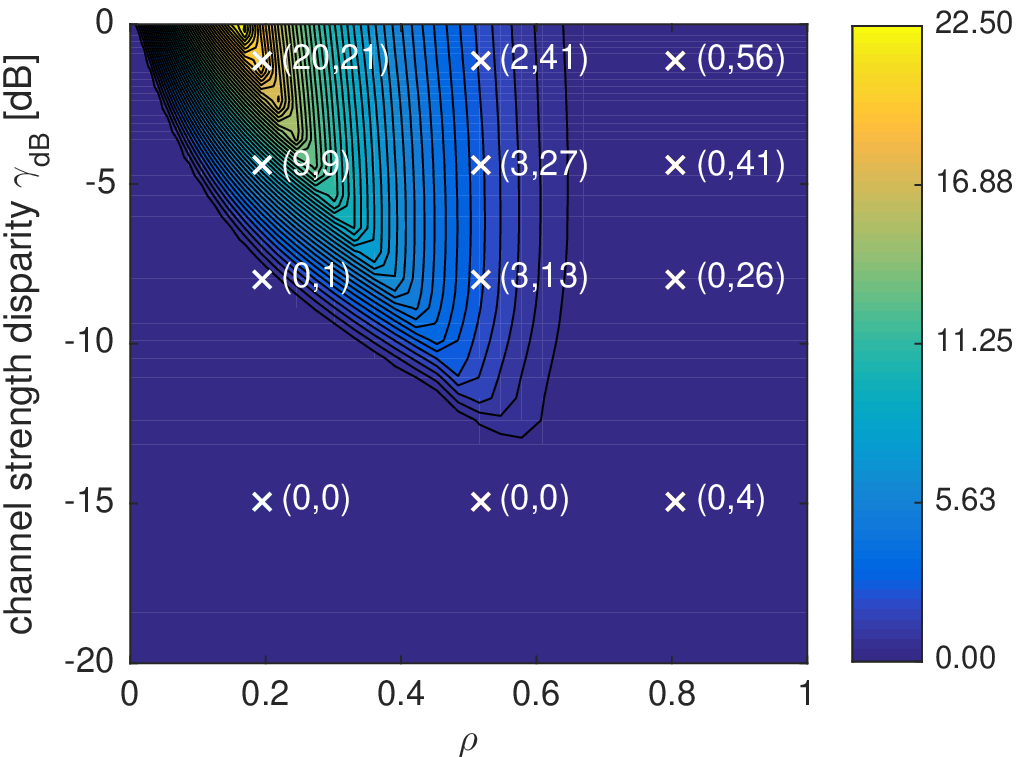}}
\subfigure[$u_1=1, u_2=10^{0.5}$]{\label{c}\includegraphics[width = 5.5cm]{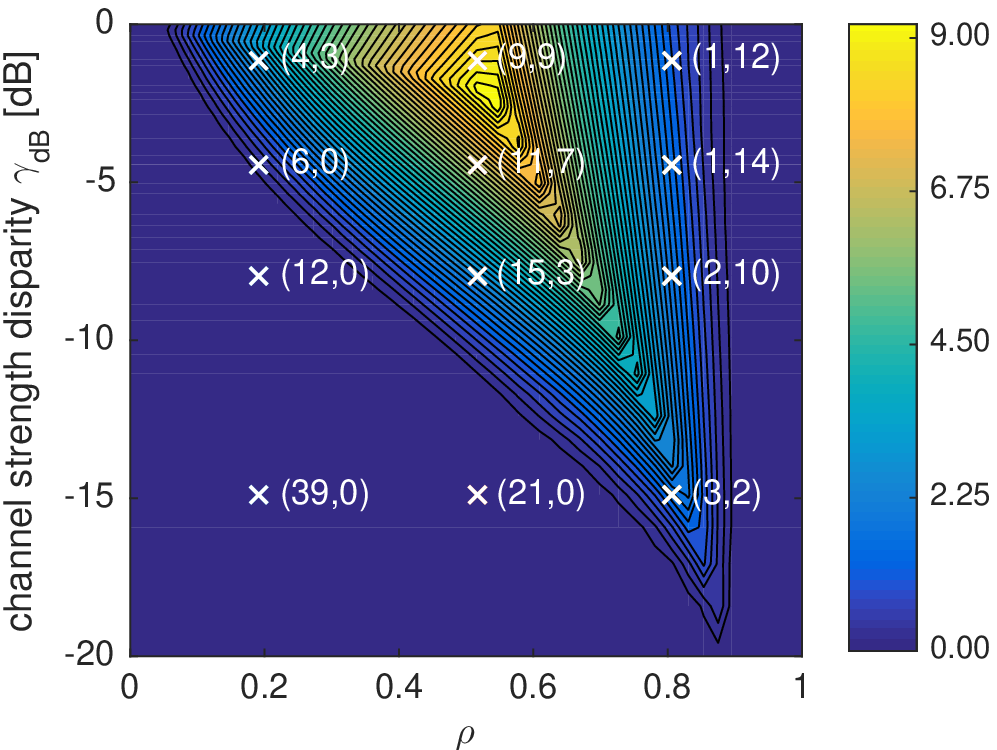}}
\end{subfigmatrix}
\vspace{-0.2cm}
\caption{Relative weighted sum-rate gain [\%] of RS over dynamic switching between SDMA and NOMA for different values of weights $u_1,u_2$, with precoders based on WMMSE optimization, $n_t=2$, and $P=100$ W. The values in brackets indicate weighted sum-rate gains over SDMA and NOMA, respectively.}
\label{gain_WMMSE}
\vspace{-0.3cm}
\end{figure*}

\begin{figure*}%[!t]
%\centering
\begin{subfigmatrix}{3}
\subfigure[$P=10$ W (SNR=10dB)]{\label{a}\includegraphics[width = 5.5cm]{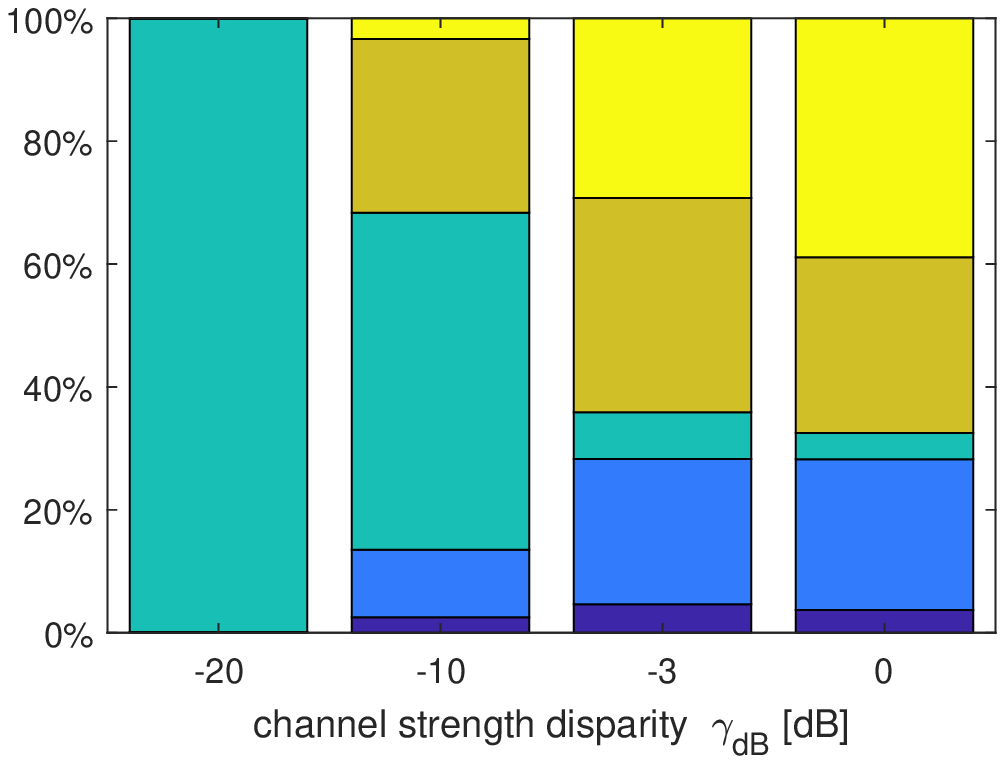}}
\subfigure[$P=100$ W (SNR=20dB)]{\label{b}\includegraphics[width = 5.5cm]{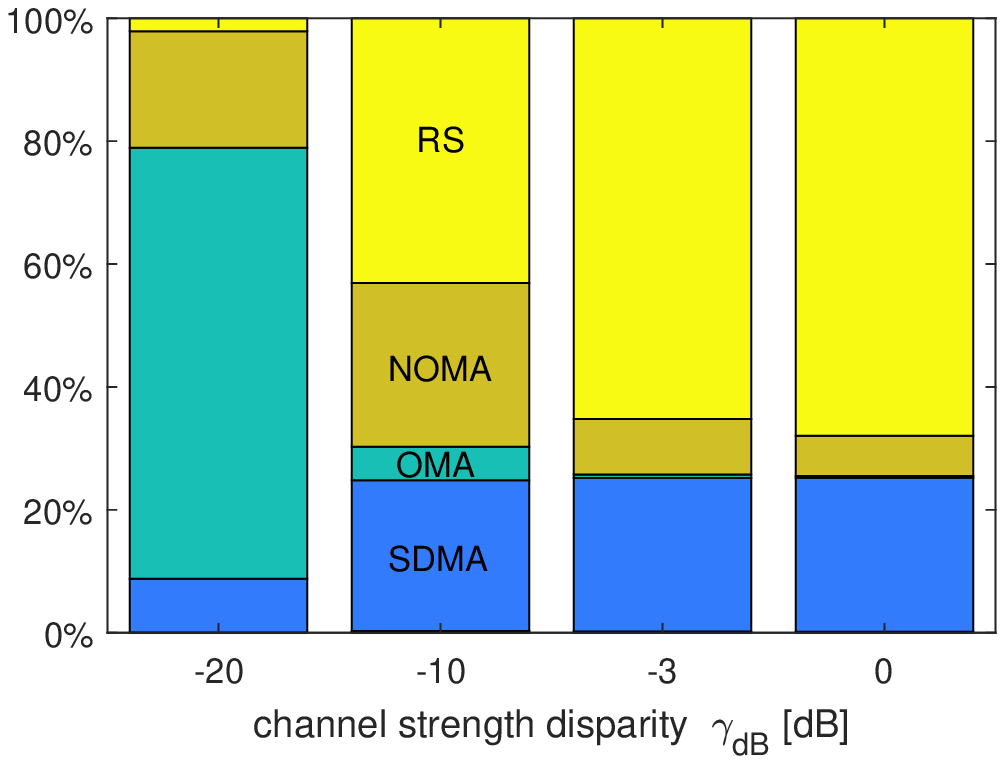}}
\subfigure[$P=1000$ W (SNR=30dB)]{\label{c}\includegraphics[width = 5.5cm]{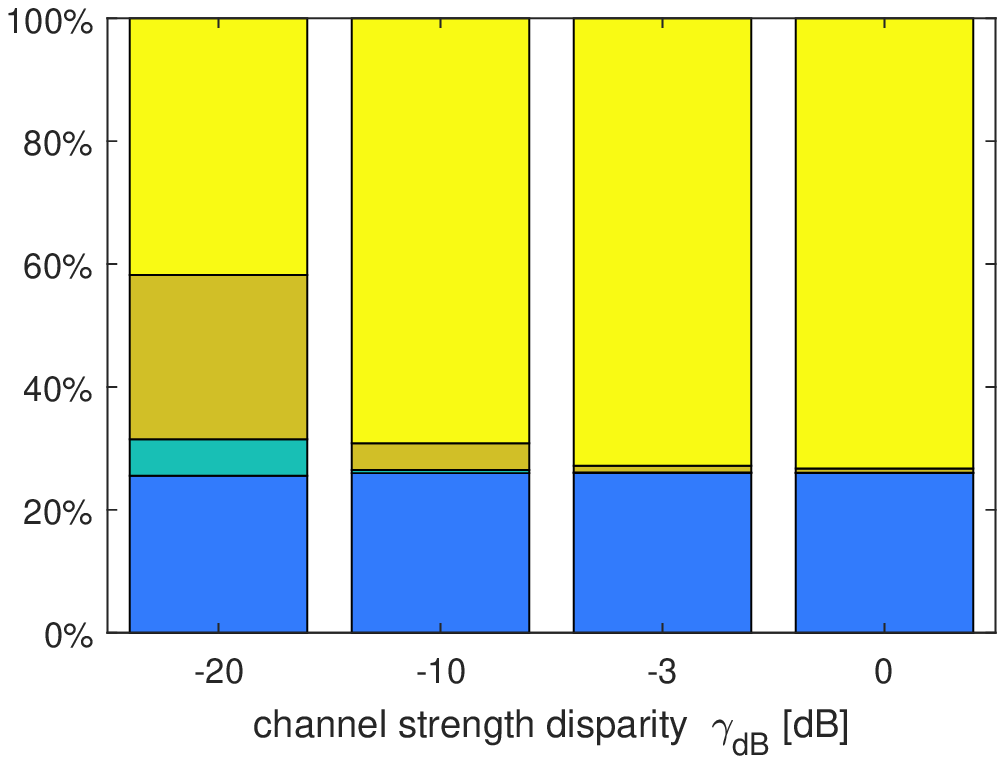}}
\end{subfigmatrix}
\vspace{-0.2cm}
\caption{Percentage of operation of RS, SDMA, NOMA, OMA, and Multicast with precoders from Section \ref{precoder_section} for $P=10 \textnormal{W}, 100 \textnormal{W}, 1000 \textnormal{W}$, with $n_t=2$.}
\label{percentage_nt2}
\vspace{-0.5cm}
\end{figure*}

%\begin{figure*}%[!t]
%%\centering
%\begin{subfigmatrix}{3}
%\subfigure[$P=10$ W (SNR=10dB)]{\label{a}\includegraphics[width = 5.5cm]{percentage_P10_nt4.eps}}
%\subfigure[$P=100$ W (SNR=20dB)]{\label{b}\includegraphics[width = 5.5cm]{percentage_P100_nt4.eps}}
%\subfigure[$P=1000$ W (SNR=30dB)]{\label{c}\includegraphics[width = 5.5cm]{percentage_P1000_nt4.eps}}
%\end{subfigmatrix}
%\caption{Percentage of operation of RS, SDMA, NOMA, OMA, and Multicast with precoders from Section \ref{precoder_section} for $P=10 \textnormal{W}, 100 \textnormal{W}, 1000 \textnormal{W}$, with $n_t=4$.}
%\label{percentage_nt4}
%\vspace{-0.3cm}
%\end{figure*}

In this section, we first illustrate the above analysis and the preferred regions for the operation of NOMA, OMA, SDMA, and RS. We assume $n_t=2$, and channel vectors given by $\mathbf{h}_1=1/\sqrt{2}\:[1,1]^H$ and $\mathbf{h}_2=\gamma/\sqrt{2}\:[1,e^{j \theta}]^H$. 

\par Assuming the precoding strategies in Section \ref{precoder_section} and the WF power allocation \eqref{WF_solution}, the colors in Fig. \ref{topt}(a) and (b) illustrate the optimum value (obtained from exhaustive search whenever not available in closed form) of $t$ that maximizes the sum-rate and the corresponding preferred communication strategy (RS, SDMA, NOMA, OMA) as a function of $\rho=1-\left|\bar{\mathbf{h}}_1^H\bar{\mathbf{h}}_2\right|^2$ (ranging from 0 to 1) and $\gamma_{\mathrm{dB}}=20\log_{10}(\gamma)$ (ranging from 0 to -20dB), i.e., user-1 and user-2 have a long-term SNR of 20dB and $0 \mathrm{dB}\leq 20 \mathrm{dB}+\gamma_{\mathrm{dB}}\leq 20 \mathrm{dB}$, respectively. Recall that SDMA is characterized by $t=1,P_1>0,P_2>0$, NOMA by $0<t<1,P_1>0,P_2\!=\!0$, OMA by $t\!=\!1,P_1\!=\!P,P_2\!=\!0$, and multicast by $t\!=\!0,P_1\!=\!0,P_2\!=\!0$. For all other regimes, RS does not specialize to any other well-established scheme and is simply referred to as RS. We observe that NOMA is preferred for deployments with small $\rho$, i.e., closely aligned users, and small $\gamma$, SDMA is preferred whenever $\rho$ is sufficiently large, i.e., semi-orthogonal users, and RS bridges those two extremes. OMA is preferred whenever $\gamma$ is very small.

\par Recall that Fig. \ref{topt} is obtained for $P=100$ W. In Fig. \ref{regions_P}, we assess the evolution of the regions as a function of $P$ for $P=10$ W and $P=1000$ W  (where the long term SNR is 10 dB and 30 dB, respectively). As $P$ increases, RS becomes the dominant strategy for most deployment conditions.

\par Fig. \ref{gain} shows the relative sum-rate gain [\%] of RS over dynamic switching between SDMA and NOMA, defined as $\frac{R_{\mathrm{s}}^{\mathrm{RS}}-\max(R_{\mathrm{s}}^{\mathrm{SDMA}},R_{\mathrm{s}}^{\mathrm{NOMA}})}{\max(R_{\mathrm{s}}^{\mathrm{SDMA}},R_{\mathrm{s}}^{\mathrm{NOMA}})}\!\times\!100$,\! for $P\!=\!10,100,1000$ W and the precoders from Section \ref{precoder_section}. RS provides explicit gains over dynamic switching for medium values of $\rho$. The values in brackets indicate the relative sum-rate gains over SDMA and NOMA, respectively, i.e., $\big(\frac{R_{\mathrm{s}}^{\mathrm{RS}}-R_{\mathrm{s}}^{\mathrm{SDMA}}}{R_{\mathrm{s}}^{\mathrm{SDMA}}}\!\times\! 100,\frac{R_{\mathrm{s}}^{\mathrm{RS}}-R_{\mathrm{s}}^{\mathrm{NOMA}}}{R_{\mathrm{s}}^{\mathrm{NOMA}}}\!\times\! 100\big)$. Large gains over SDMA are observed for low to medium values of $\rho$, and over NOMA for medium to large values of $\rho$ at low SNR and for all values of $\rho$ and $\gamma_{\mathrm{dB}}$ at higher SNR. Values $(0,0)$ indicate that OMA is the preferred strategy, and that RS, SDMA, and NOMA all specialize to OMA.

%\par In Fig. \ref{high_SNR}, we consider the same channel model as in Fig. \ref{topt}/\ref{regions} and illustrate the behavior of $t^{\star}$ in \eqref{optt_highsnr} and $\Delta R_{\mathrm{s}}$ in \eqref{delta_Rs}.  
%
%\begin{figure}%[!t]
%\centering
%\includegraphics[width=0.8\columnwidth]{high_SNR_t_delta.eps}
%\caption{Optimum $t$ and sum-rate gap between RS and SDMA at high SNR.}
%\label{high_SNR}
%\vspace{-0.3cm}
%\end{figure}
%\vspace{-0.3cm}

\par Fig. \ref{gain_WMMSE} is similar to Fig. \ref{gain} but now the Weighted Minimum Mean Square Error (WMMSE) precoding optimization framework for RS developed in \cite{Joudeh:2016a,Mao:2017} is adopted. Such framework optimizes all precoders ($\mathbf{p}_{\mathrm{c}}, \mathbf{p}_1, \mathbf{p}_2$) jointly with the power allocations so as to maximize the weighted sum-rate $\sum_{k=1,2} u_k \left(R_k+R_{\mathrm{c},k}\right)$. In those evaluations, the convergence tolerance of the WMMSE algorithm is set to $\epsilon\!=\!10^{-3}$ \cite{Mao:2017}. %When deciding the regions of the different strategies, the optimized results of $t$, $P_1$ and $P_2$ are rounded to 1 decimal place. 
When allocating equal weights or higher weights to the user with the stronger channel (namely user-1), NOMA has no benefit over SDMA. When a higher weight is given to the weaker user (user-2), NOMA is able to outperform SDMA. RS on the other hand always provides the same or better performance than both SDMA and NOMA for all weights, $\rho$, and $\gamma_{\mathrm{dB}}$. Though the precoders of Section \ref{precoder_section} are simple and not optimal, the insights obtained from the analysis and Fig. \ref{gain} are inline with those obtained from Fig. \ref{gain_WMMSE}. Hence, irrespectively of the precoding strategies, i.e., simple or optimized, RS unifies and outperforms SDMA, OMA, NOMA, and multicasting.

\par We now change the channel model and assume i.i.d. Rayleigh fading, i.e., the entries of $\mathbf{h}_1$ and $\mathbf{h}_2$ are $\mathcal{CN}(0,1/n_t)$ and $\mathcal{CN}(0,\gamma^2/n_t)$. We generate 10000 channel realizations. Making use of the precoders in Section \ref{precoder_section}, we identify the preferred (i.e., sum-rate maximizing) strategy for each channel realization. Fig. \ref{percentage_nt2} displays the percentage a given strategy is the preferred option as a function of $P$ and $\gamma_{\mathrm{dB}}$ for $n_t=2$. OMA is preferred for low $P$ and low $\gamma_{\mathrm{dB}}$, and RS becomes the preferred option as $P$ and/or $\gamma_{\mathrm{dB}}$ increase. At high SNR, RS is the preferred option for about 75\% of the channel realizations and SDMA for the remaining 25\%. Results with $n_t=4$ (not reproduced here due to the space constraint) show that NOMA almost disappears from the set of preferred strategies, and SDMA becomes more dominant (for about 60\% of the channel realizations and RS for the remaining 40\%). This is natural since, as $n_t$ increases, the likelihood to experience large $\rho$ increases, and $t^{\star}$ has a higher chance of being equal to 1. 
\vspace{-0.3cm}
\section{Conclusions}\label{conclusions}
RS unifies SDMA, OMA, NOMA, and multicasting under a single approach and provides a powerful framework for the design and optimization of non-orthogonal transmission, multiple access, and interference management strategies. Thanks to its versatility, RS has the potential to tackle challenges of modern communication systems and is a gold mine of research problems for academia and industry, spanning fundamental limits, optimization, PHY and MAC layers, and standardization. 
\vspace{-0.5cm}

\ifCLASSOPTIONcaptionsoff
  \newpage
\fi

\end{document}